\documentclass[prl,reprint,superscriptaddress,nofootinbib]{revtex4-1}  %% LaTeX 2e (preferred)
\pdfoutput=1
\usepackage{amssymb,amsmath}
\usepackage{graphicx}
\usepackage{natbib}
\usepackage{mathrsfs}
\usepackage{multibib}
\usepackage[pdftex=true,linktoc=page]{hyperref}
\usepackage[T1]{fontenc}
\usepackage[letterpaper,textwidth=7in,top=.75in,bottom=.75in]{geometry}
\newcommand{\ket}[1]{|#1\rangle}

\begin{document}
\title{Dynamic stabilization of the optical resonances of single nitrogen-vacancy centers in diamond}

 \author{V.~M.~Acosta}
    \email{victor.acosta@hp.com}
    \address{
     Hewlett-Packard Laboratories, 1501 Page Mill Rd.,
     Palo Alto, CA 94304
    }
 \author{C.~Santori}
    \address{
     Hewlett-Packard Laboratories, 1501 Page Mill Rd.,
     Palo Alto, CA 94304
    }
  \author{A.~Faraon}
    \address{
     Hewlett-Packard Laboratories, 1501 Page Mill Rd.,
     Palo Alto, CA 94304
    }
 \author{Z.~Huang}
    \address{
     Hewlett-Packard Laboratories, 1501 Page Mill Rd.,
     Palo Alto, CA 94304
    }
  \author{K.-M.~C.~Fu}
    \address{
     Hewlett-Packard Laboratories, 1501 Page Mill Rd.,
     Palo Alto, CA 94304
    }
    \address{
    Departments of Physics and Electrical Engineering, University of Washington,
    Seattle, WA 98195
    }
  \author{A.~Stacey}
    \address{
     School of Physics, University of Melbourne,
    Melbourne VIC 3010, Australia
    }
  \author{D.~A.~Simpson}
    \address{
     School of Physics, University of Melbourne,
     Melbourne VIC 3010, Australia
    }
  \author{K.~Ganesan}
    \address{
     School of Physics, University of Melbourne,
     Melbourne VIC 3010, Australia
    }
  \author{S.~Tomljenovic-Hanic}
    \address{
     School of Physics, University of Melbourne,
     Melbourne VIC 3010, Australia
    }
  \author{A.~D.~Greentree}
    \address{
     School of Physics, University of Melbourne,
     Melbourne VIC 3010, Australia
    }
    \address{Applied Physics, School of Applied Sciences, RMIT University, Melbourne 3001, Australia}
  \author{S.~Prawer}
    \address{
     School of Physics, University of Melbourne,
     Melbourne VIC 3010, Australia
    }
 \author{R.~G.~Beausoleil}
    \address{
     Hewlett-Packard Laboratories, 1501 Page Mill Rd.,
     Palo Alto, CA 94304
    }
%\date{\today}

\begin{abstract}
We report electrical tuning by the Stark effect of the excited-state structure of single nitrogen-vacancy (NV) centers located $\lesssim100~{\rm nm}$ from the diamond surface. The zero-phonon line (ZPL) emission frequency is controllably varied over a range of $300~{\rm GHz}$.  Using high-resolution emission spectroscopy, we observe electrical tuning of the strengths of both cycling and spin-altering transitions. Under resonant excitation, we apply dynamic feedback to stabilize the ZPL frequency. The transition is locked over several minutes and drifts of the peak position on timescales $\gtrsim100~{\rm ms}$ are reduced to a fraction of the single-scan linewidth, with standard deviation as low as 16 MHz (obtained for an NV in bulk, ultra-pure diamond). These techniques should improve the entanglement success probability in quantum communications protocols.
%These techniques can increase the success probability of photon-heralded NV-NV entanglement.

%07.55.Ge Magnetometers for magnetic field measurements, 61.72.jn Color centers
%76.30.Mi Color centers and other defects
%81.05.Uw  Carbon, diamond, graphite
\end{abstract}
%\pacs{(07.55.Ge) Magnetometers for magnetic field measurements; (61.72.jn) Color centers; (76.30.Mi) Color centers and other defects; (81.05.Uw) Carbon, diamond, graphite}
\maketitle

Integrated photonic networks based on cavity-coupled solid-state spin impurities offer a promising platform for scalable quantum computing \cite{OBR2007,STE2008,BEN2009,LAD2010,SAN2010}. A key ingredient for this technology is the generation and interference of indistinguishable photons emitted by pairs of identical spin qubits \cite{CAB1999,CHI2005,MOE2007}. This requires spectrally stable emitters with identical level structure, a formidable challenge in the solid-state environment.

A potential solution is to use external control to counteract sample inhomogeneities. In candidate systems based on single molecules \cite{ORR1992,WIL1992,LET2010}, quantum dots \cite{EMP1997,PAT2010}, and negatively-charged nitrogen-vacancy (NV) centers in diamond \cite{TAM2006,BAS2011,BER2012}, the level structure can be statically tuned via the DC Stark effect. However, the spectral stability of emitters in these systems is often hampered by local fluctuations which cause the emission frequency to change with time, a phenomenon known as spectral diffusion \cite{AMB1991}.  Previous attempts to address this problem have focused on improving the host material \cite{JEL1997,SAN2002,TAM2006,GRE2006} or using post-selection techniques \cite{ATE2009,ROB2010,TOG2010,BER2012}, but a robust, high-yield solution is still lacking.

The diamond NV center is an attractive spin qubit, as it exhibits a unique combination of long-lived spin coherence \cite{BAL2009} and efficient optical control and readout \cite{BUC2010,ROB2011}. However integration into on-chip photonic networks requires NV centers to be located near nanostructured surfaces, where inhomogenous strain and spectral diffusion can be particularly problematic \cite{FU2010,FAR2011}. In this Letter, we first demonstrate electrical control over the zero-phonon line (ZPL) transition frequencies, as well as probabilities for both cycling and $\Lambda$-type transitions, of single NV centers located near the diamond surface. We then show that spectral diffusion of the ZPL can be suppressed to 16 MHz standard deviation, on timescales $\gtrsim100~{\rm ms}$, by providing rapid electrical feedback to compensate for local field fluctuations.

\begin{figure}
\centering
    \includegraphics[width=.48\textwidth]{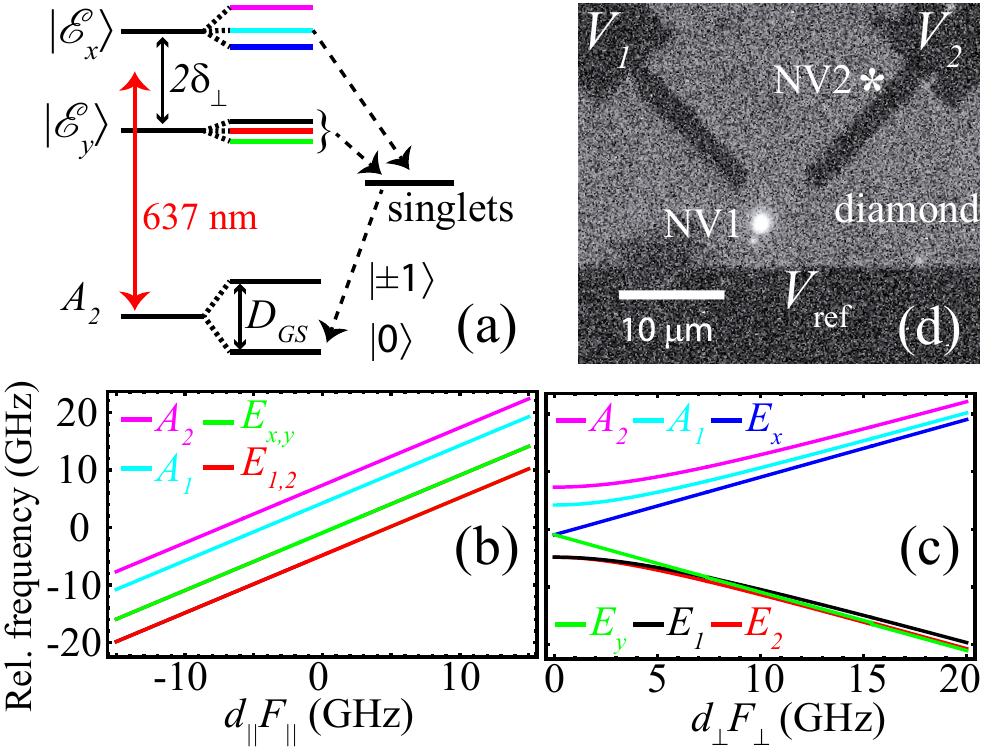}
    \caption{\label{fig:appStark} (a) NV level structure at high transverse field ($d_{\perp} F_{\perp}\equiv\delta_{\perp}>15~{\rm GHz}$). Dashed lines indicate the spin-selective decay path responsible for optical pumping. (b) Effect of longitudinal and (c) transverse electric fields on the excited state levels. (d) Fluorescence micrograph of the electrode structure. NV1 was illuminated by 532 nm light and appears white, while the position of NV2 is denoted with an asterisk. Metal electrodes appear as shadows.
    }
    %, and the slight discontinuities correspond to avoided level crossings.
\end{figure}
 The negatively-charged nitrogen-vacancy (NV) center has $C_{3v}$ symmetry, and the basic energy structure is depicted in Fig. \ref{fig:appStark}(a). The spin-triplet ground state, ${^3}A_2$, is split such that the $m_s=0$ spin projection (labeled $\ket{0}$ throughout) is separated from the degenerate $m_s=\pm1$ levels ($\ket{\pm1}$) by $D_{GS}=2.88~{\rm GHz}$ at $T\lesssim100~{\rm K}$ \cite{ACO2010}. The optically excited state (ES) is a spin triplet and orbital doublet with ${^3}E$ symmetry, and its fine structure has been studied theoretically in detail \cite{DOH2011,MAZ2011}. The Hamiltonian describing the ES manifold is:
%Briefly, a combination of spin-orbit and spin-spin interaction dominate in the absence of external electric and/or strain fields, resulting in a basis set of six symmetry states, $\{\ket{A_2},\ket{A_1},\ket{E_x},\ket{E_y},\ket{E_1},\ket{E_2}\}$, ordered from highest to lowest energy.
\begin{align}
\label{eq:hamiltonian}
&\mathscr{H}_{ES}=\mathscr{H}_{SO}+\mathscr{H}_{SS}+\mathscr{H}_{Stark},\nonumber
\\
&\mathscr{H}_{Stark}=-\vec{d}\cdot\vec{F},
\end{align}
where $\mathscr{H}_{SO}$, $\mathscr{H}_{SS}$, and $\mathscr{H}_{Stark}$ contain respectively, the spin-orbit, spin-spin, and Stark effect contributions; $\vec{d}$ is the electric dipole moment, and $\vec{F}$ is the electric field. The effect of a strain is treated as an effective electric field \cite{HUG1967,DAV1976}.

We first consider the influence of $\mathscr{H}_{Stark}$ on only the orbital portion of the ES wavefunction, consisting of two eigenstates, $\{\ket{\mathscr{E}_{x}},\ket{\mathscr{E}_{y}\}}$, initially degenerate at zero field.
%These states have $m_s=0$ spin-projection and consequently commute with $\mathscr{H}_{SO}$, while the effect of $\mathscr{H}_{SS}$ on these energy levels is order $0.1~{\rm GHz}$ and neglected for now.
Under electric fields, the orbitals exhibit energy shifts, $\Delta(\vec{F})$, of:
\begin{align}
\label{eq:exy}
\Delta_{\mathscr{E}_{x}}(\vec{F})&=d_{\parallel} F_{\parallel}+d_{\perp} F_{\perp},\nonumber \\
\Delta_{\mathscr{E}_{y}}(\vec{F})&=d_{\parallel} F_{\parallel}-d_{\perp} F_{\perp},
\end{align}
where the directions are with respect to the NV symmetry axis. Longitudinal fields do not lift the orbital degeneracy and result only in equal, linear shifts of all levels. Transverse fields split the orbitals into two branches with an energy difference, $2\delta_{\perp}\equiv(\Delta_{\mathscr{E}_{x}}-\Delta_{\mathscr{E}_{y}})$, that grows linearly with increasing field. The spacings between ground-state sublevels remain relatively unaffected by electric fields \cite{VAN1990,DOL2011}. The ground state may have a longitudinal dipole moment, $d_{GS,\parallel}$ \cite{MAZ2011,DOH2011}, but in experiments we only resolved $\Delta d_{\parallel}\equiv d_{\parallel}-d_{GS,\parallel}$,

Incorporating spin interactions results in a set of six eigenstates, $\{\ket{A_2},\ket{A_1},\ket{E_x},\ket{E_y},\ket{E_1},\ket{E_2}\}$, ordered from highest to lowest energy (at low field). Figures \ref{fig:appStark}(b) and (c) show the effect of $\mathscr{H}_{ES}$ on all six ES energies due to, respectively, $F_{\parallel}$ and $F_{\perp}$. %As will be discussed, an important feature is the presence of transverse-field-dependent avoided crossings amongst levels in the lower orbital branch, comprised of $\{\ket{E_y},\ket{E_1},\ket{E_2}\}$.

We focused most of our study on NV centers close to the diamond surface, a necessary feature for future integration with nanophotonic devices. Our sample, described in detail elsewhere \cite{STA2012APL,STA2012}, consists of a high-purity single-crystal, [100]-oriented diamond substrate with a $\sim 100~{\rm nm}$ thick chemical-vapor-deposition-grown layer with [NV]$\approx10^{6}~{\rm cm^{-2}}$. %The top $\sim50~{\rm \mu m}$ of the substrate has been plasma treated \cite{STA2011ARXIV} so as to exhibit virtually no NV fluorescence. An approximately $100~{\rm nm}$ layer of diamond was grown on top of this substrate and the resulting sheet density of NV$^{\mbox{-}}$ centers is $\sim10^{6}~{\rm cm^{-2}}$.
%An approximately $100~{\rm nm}$ layer of diamond was grown on top of this substrate.
The two NV centers studied in this work, labeled NV1 and NV2, are located in this surface layer \cite{STA2012}. Lithographically-defined metal electrodes \cite{BAS2011} were deposited on the surface [Fig. \ref{fig:appStark}(d)]. The layout of the electrodes (labeled $V_1, V_2$ and $V_{\mathrm{ref}}$) permits tuning of electric fields in any in-plane direction near the center of the structure [Supplementary Information].
%Application of $10~{\rm V}$ to one electrode (with the other two electrodes grounded) corresponds to an electric field amplitude at the location of NV1 of $0.9, 0.6,$ and $1.1~{\rm MV/m}$, for $V_1, V_2,$ and $V_{ref}$, respectively, as determined from electrostatic modeling treating the diamond as a perfect dielectric (Supplementary Information).

A confocal microscope was used to excite and collect emitted light from a diamond sample in thermal contact with the cold finger of a flow-through, liquid-helium cryostat. The cold finger was maintained at a temperature $T\approx7~{\rm K}$, and no magnetic field was applied. Two forms of spectroscopy were employed: high-resolution emission spectroscopy and photoluminescence excitation spectroscopy.

\begin{figure}
\centering
    \includegraphics[width=.47\textwidth]{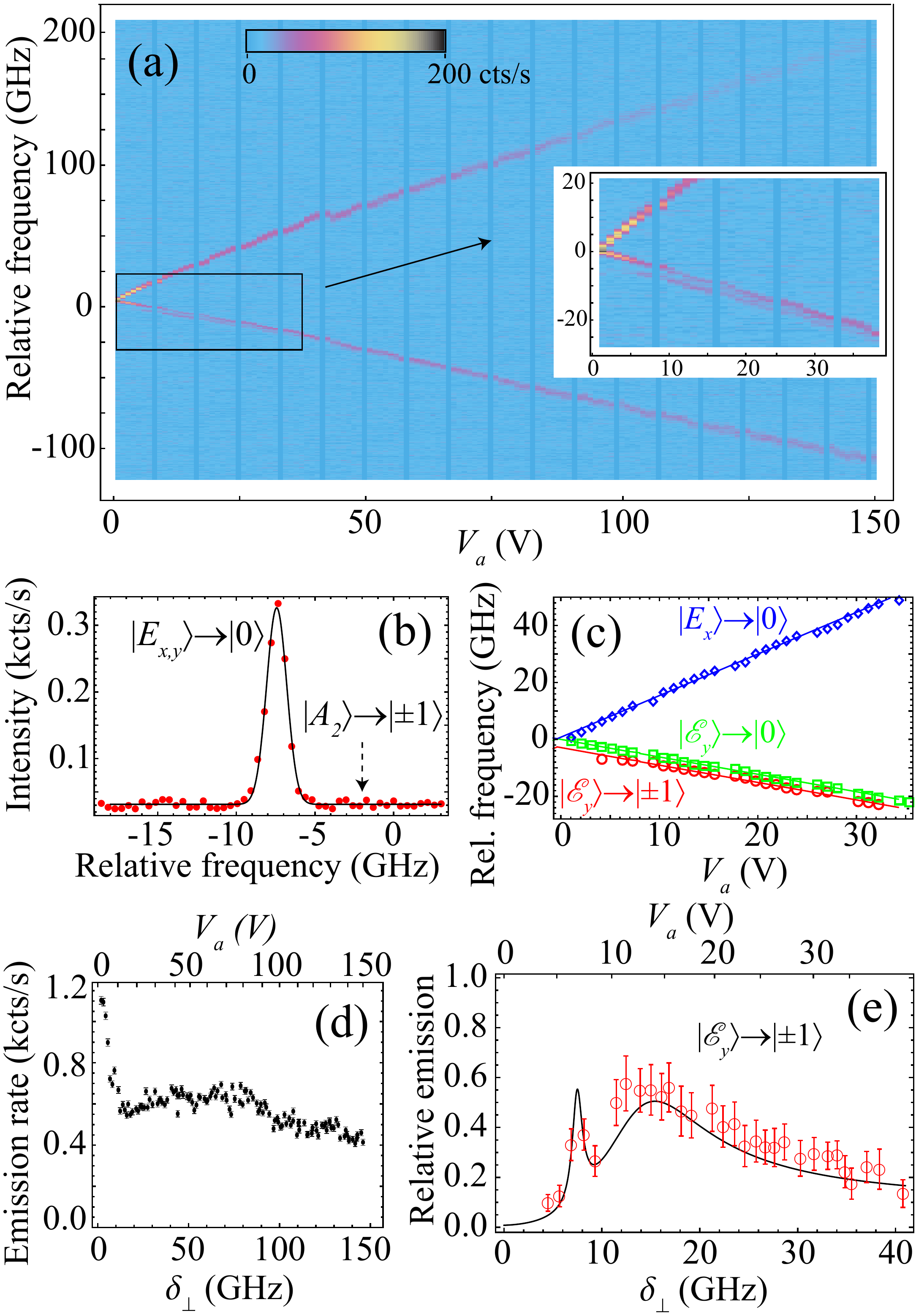}
    \caption{\label{fig:emission} (a) Stark emission spectroscopy. $V_a$ was slowly scanned ($1~{\rm V/min}$) and emission spectra were obtained in one-minute intervals. A dark exposure frame was taken every 8th frame (cyan vertical stripes). Emission frequencies are relative to $470.45~{\rm THz}~ (637.25~{\rm nm})$. Low-field data are inset. (b) Emission spectrum for $V_a=1~{\rm V}$ with expected peak positions labeled. The $\ket{E_{x,y}}\rightarrow\ket{0}$ emission line is fit with a Gaussian profile. (c) Low-field peak positions from (a) and global fit based on Eq. \eqref{eq:hamiltonian}. Lorentzian fit uncertainty is smaller than the plotted symbols. (d) Total ZPL emission versus $\delta_{\perp}$. The emission rate was calculated by subtracting the mean background and summing counts over a range of $\sim3$ FWHM linewidths centered at each peak. Error bars are based on Poissonian noise. (e) Relative intensity of the $\ket{\mathscr{E}_y}\rightarrow\ket{\pm1}$ emission line along with fit (see text).
    %(f) Relative intensity of emission from $\ket{E_x}$ and $\ket{E_y}$ branches computed as in (d). We eliminate the effect of unequal excitation and/or polarization-dependent scattering in our detection optics by normalizing the emission intensities of each branch to their 0-field values. The intensity for $\ket{E_y}$ branch includes both emission lines, when visible. The fits are to one and two-phonon orbital relaxation using the model in (g), as described in text.
    }
\end{figure}

For emission spectroscopy, $2~{\rm mW}$ of 532-nm laser light was focused by a 0.6-numerical-aperture objective onto NV1, exciting through the phonon sideband (PSB) near saturation. The collected emission was spectrally filtered to direct ZPL light ($636\mbox{-}638~{\rm nm}$) to a high-resolution grating spectrometer. The optical polarization was chosen to ensure excitation of both orbital branches \cite{FU2009}. Figure \ref{fig:emission}(a) shows the emission spectra versus voltage, $V_a$, applied simultaneously on $V_1$ and $V_{\mathrm{ref}}$, with $V_2=0$. By varying $V_a$ from $0$ to $150~{\rm V}$, we observe linear tuning of emission lines over a range exceeding $300~{\rm GHz}$. Such a wide tuning range, enabled by the enhanced fields provided by our devices [Supplementary Information], is essential to compensate for the large intrinsic fields typical in nanophotonic devices \cite{FAR2011}.

Depending on the applied voltage, we resolve between one and three emission lines. At $V_a\approx1~{\rm V}$ we observe a single emission line [Fig. \ref{fig:emission}(b)] with full-width-at-half-maximum (FWHM) linewidth of $1.4(2)~{\rm GHz}$, near the spectrometer resolution of $\sim0.9~{\rm GHz}$. We interpret this peak as containing unresolved contributions from the $\ket{E_{x,y}}\rightarrow\ket{0}$ cycling transitions \cite{TAM2008,BAT2009}. Taking into consideration the absence of other peaks, in particular the $\ket{A_2}\rightarrow\ket{\pm1}$ cycling transition \cite{TOG2010}, and the observed noise floor, we place a bound on the ground-state spin polarization $\mathscr{P}_{GS}\equiv P_0/(P_0+P_{\pm1})\gtrsim90\%$, where $P_i$ is the occupation probability of state $\ket{i}$.

%In the absence of external fields, the $\ket{A_2}, \ket{A_1}, \ket{E_1},$ and $\ket{E_2}$ levels contain linear combinations of $m_s=\pm1$ spin states, while the $\ket{E_x}$ and $\ket{E_y}$ levels contain primarily $m_s=0$.
Upon application of transverse fields, the spin character of the levels in the upper $\ket{\mathscr{E}_{x}}$ orbital branch, $\{\ket{A_2},\ket{A_1},\ket{E_x}\}$, remain relatively unperturbed.
%Consequently, regardless of local strain, $\ket{E_x}\leftrightarrow\ket{0}$ and $\ket{A_2}\leftrightarrow\ket{\pm1}$ form nearly-closed cycling transitions \cite{TOG2010}, allowing hundreds of optical cycles without altering spin projection \footnote{The $\ket{A_1}\leftrightarrow\ket{\pm1}$ is not a good cycling transition due to strong coupling between $\ket{A_1}$ and the singlet levels \cite{TOG2010}.}.
In contrast, the spin character of levels in $\ket{\mathscr{E}_{y}}$, $\{\ket{E_y},\ket{E_1},\ket{E_2}\}$, mix at avoided crossings due to spin-spin interaction \cite{SAN2006OPTEX,TAM2008,DOH2011,MAZ2011}, making these levels useful for spin-altering $\Lambda$ schemes.

In the range $5~{\rm V}\lesssim V_a\lesssim35~{\rm V}$, three emission lines are visible [inset of Fig. \ref{fig:emission}(a)]. Based on the positive, linear tuning, the upper peak is identified as $\ket{E_x}\rightarrow\ket{0}$. Lorentzian fits to these spectra reveal that the two lowest lines are on average separated by $2.7(2)~{\rm GHz}$, which is comparable to $D_{GS}$. Considering also the negative, linear tuning, we conclude that these peaks arise from $\ket{\mathscr{E}_{y}}\rightarrow\ket{0}$ and $\ket{\mathscr{E}_{y}}\rightarrow\ket{\pm1}$ emission (the three levels within $\ket{\mathscr{E}_{y}}$ are nearly degenerate and unresolved here).  The presence of these lines was previously predicted based on observations of spin-altering $\Lambda$-type transitions involving the lower orbital branch \cite{SAN2006OPTEX,SAN2006,TAM2008}. Figure \ref{fig:emission}(c) plots the emission frequencies along with a fit using a model based on Eq. \eqref{eq:hamiltonian}, showing excellent agreement. The fitted parameters are $\Delta d_{\parallel} F_{\parallel}/V_a=0.42(2)~{\rm GHz/V}$ and $d_{\perp} F_{\perp}/V_a=1.03(3)~{\rm GHz/V}$.

Even with significant emission to $\ket{\pm1}$, we still do not observe $\ket{A_2}\rightarrow\ket{\pm1}$ emission. Throughout, we find $\mathscr{P}_{GS}\gtrsim85\%$. A likely explanation is that any population in $\ket{\pm1}$ is quickly transferred to the metastable singlet levels \cite{ACO2010PRB}, preventing the detection of $m_s=\pm1$ emission lines. This is consistent with Fig. \ref{fig:emission}(d), where the total ZPL emission rate integrated over all lines is plotted as a function of one half the orbital splitting, $\delta_{\perp}$. Between $3\lesssim\delta_{\perp}\lesssim10~{\rm GHz}$ ($3\lesssim V_a\lesssim10~{\rm V}$), the emission rate falls precipitously before leveling off at less than half the initial rate.

The relative intensity of the emission lines gives further insight into the ES properties. Figure \ref{fig:emission}(e) plots the intensity of the $\ket{\mathscr{E}_{y}}\rightarrow\ket{\pm1}$ emission line, normalized by the total emission from $\ket{\mathscr{E}_y}$, as a function of $\delta_{\perp}$. Evidently, the applied field is a powerful knob in tuning the relative transition strengths in this $\Lambda$ system. Two peaks for the emission of $\ket{\mathscr{E}_{y}}\rightarrow\ket{\pm1}$ are present at $\delta_{\perp}\approx7$ and $15~{\rm GHz}$. These features correspond to level anticrossings [see Fig. \ref{fig:appStark}(c)], where maximal mixing of levels in the lower orbital branch occurs. The degree of mixing depends sensitively on both the magnitude of the transverse electric field and its angle, $\theta_r$, with respect to the $C_{3v}$ reflection planes \cite{TAM2008}. We model the relative emission intensity by assuming the NV center is excited from $\ket{0}$ to one of the three levels in the lower branch, $\ket{\mathscr{E}_{y,i}}$. The probability that emission is back to
%$\ket{0}$ is then $\sum_i P_{0,i}(\theta_r,\delta_{\perp})^2$, while the probability for emission back to
$\ket{\pm1}$ is then $\sum_i P_{0,i}(\theta_r,\delta_{\perp})(1-P_{0,i}(\theta_r,\delta_{\perp}))$, where $P_{0,i}(\theta_r,\delta_{\perp})$ is calculated by taking the overlap of $\ket{\mathscr{E}_{y,i}}$ with $m_s=0$ and tracing over orbital degrees of freedom. Here we assume all levels in $\ket{\mathscr{E}_{y}}$ couple equally to the singlets. Using the model based on Eq. \eqref{eq:hamiltonian}, we fit this formula to the data and find good agreement for $\theta_r=15(5)^{\circ}$. %and take the overlap of each state with $m_s=0$ and tracing over orbital degrees of freedom

%weighted by the probability of $m_s=0$ projection, $P_{0,i}(\theta_r,\delta_{\perp})=|\langle \tilde{E_x}\ket{\mathscr{E}_{y,i}}|^2+|\langle \tilde{E_y}\ket{\mathscr{E}_{y,i}}|^2$. Here $\tilde{E_{x,y}}$ are eigenstates of $\mathscr{H}_{SO}$ alone, which contain purely $m_s=0$ projection.

%The probability that emission is back to
%$\ket{0}$ is then $\sum_i P_{0,i}(\theta_r,\delta_{\perp})^2$, while the probability for emission back to
%$\ket{\pm1}$ is then $\sum_i P_{0,i}(\theta_r,\delta_{\perp})(1-P_{0,i}(\theta_r,\delta_{\perp}))$. Using the model based on Eq. \eqref{eq:hamiltonian}, we fit this formula to the data and find good agreement for $\theta_r=15(5)^{\circ}$.

Finally, we also observe a strong dependence of the relative emission between the upper and lower branches on $\delta_{\perp}$ [Fig. \ref{fig:emission}(a)]. Given the low temperature $T=7~{\rm K}$, this may be due to a single-phonon orbital relaxation process. Phonon decay to the lower branch could contribute to the decreased total emission in Fig. \ref{fig:emission}(d). A detailed study will be the focus of future work. All of the effects described above were reproduced in subsequent voltage scans [Supplementary Information]. %\textbf{The subsequent voltage scans were on different electrodes. Should I include the extra data in Supp Info or is this convincing enough already?}.

%The NV excited state is subject to dynamic Jahn-Teller distortions \cite{FU2009,ABT2011}, which, for $T\gtrsim10~{\rm K}$, induces orbital relaxation via a two phonon-process \cite{FU2009}. In Fig. \ref{fig:emission}(f), we plot the relative emission intensity from the two ES orbitals versus $\delta_{\perp}$. We notice a surprisingly strong dependence which, given the cold-finger temperature was $T\approx7~{\rm K}$, is unlikely to be due to the aforementioned two-phonon process. We consider a simple model [Fig. \ref{fig:emission}(g)] similar to Ref. \cite{FU2009} which includes Boltzmann-weighted orbital relaxation between $E_x$ and $E_y$, parameterized by the rate $\Gamma$, and overall decay to, for example, the ground state, with rate $\gamma_{dec}=(12~{\rm ns})^{-1}$. For a single phonon process, we expect $\Gamma=a_1 \delta_{\perp}^3$ due to the spectral-weighted phonon density of states \cite{FU2009} \textbf{need better ref}. For a two phonon process we expect $\Gamma=a_2 \delta_{\perp}^5$. Fits to these models are overlayed in Fig. \ref{fig:emission}(f). While the model does not describe the low-field data accurately, potentially due to orbital-dependent coupling to singlet levels, the best fit is for the single phonon model where we find $a_1=2.3(2)\times10^{-7}~{\rm GHz^{-3} ns^{-1}}$. This phenomenon may be related to the orbital dephasing observed in other low-temperature work \cite{ROB2010}.

%\begin{figure*}
\begin{figure}
\centering
    \includegraphics[width=.45\textwidth]{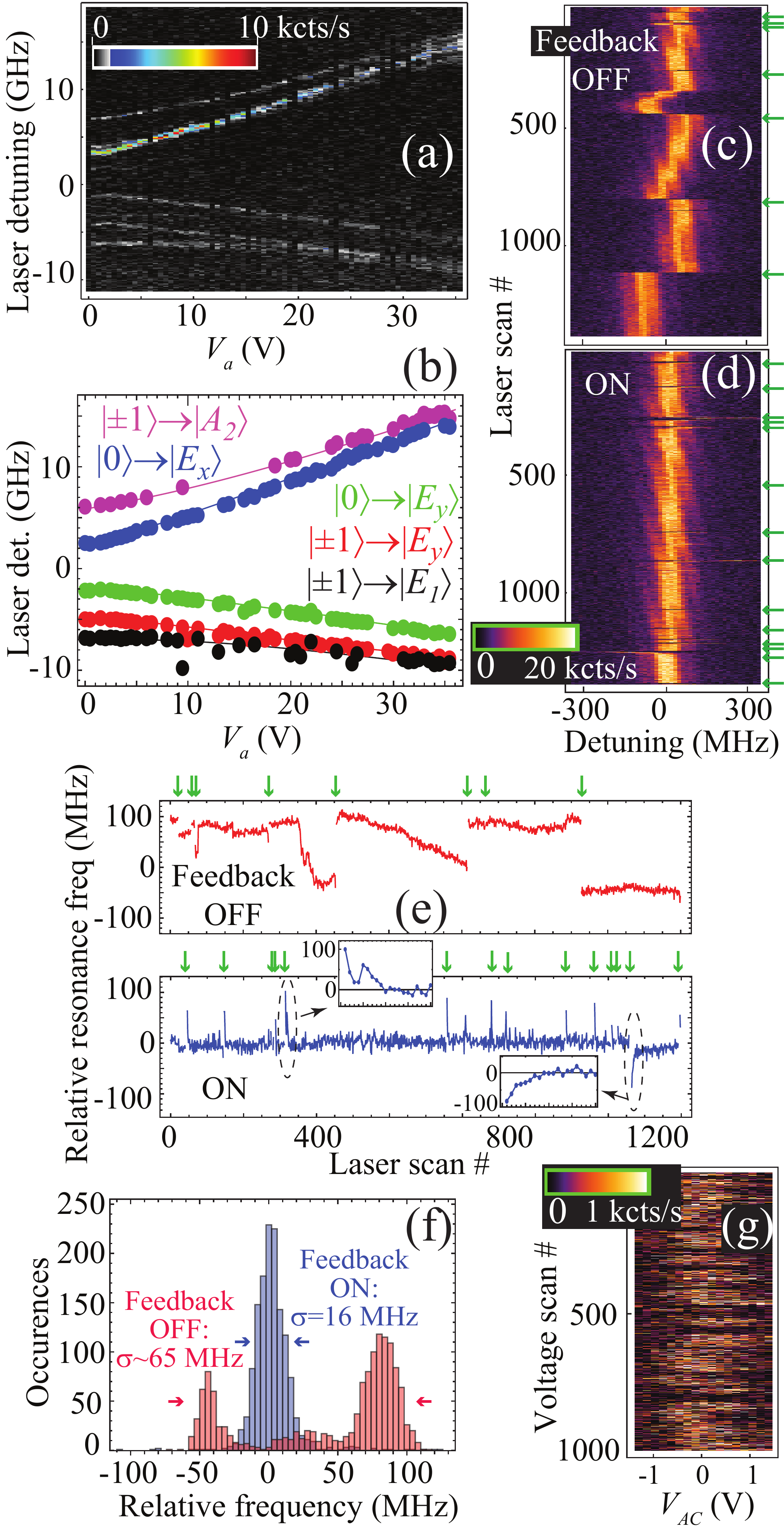}
    \caption{\label{fig:feedback} (a) PLE spectra versus applied voltage for NV1. A $532~{\rm nm}$ repump pulse ($2.5~{\rm \mu W}$, $1~{\rm s}$ duration) was applied every $60~{\rm s}$. (b) Fitted transition frequencies from (a) and global fit based on Eq. \eqref{eq:hamiltonian}, with transitions labeled. (c) PLE spectra for a single NV in the Ural sample ($\ket{0}\rightarrow\ket{E_x}$) at $V_a=-4~{\rm V}$. Green arrows indicate when the repump was applied. (d) PLE spectra for the same NV with voltage feedback applied. (e) Fitted peak positions for scans in (c) [upper panel] and (d) [lower panel]. The recovery under feedback after two separate repump pulses are inset. (f) Histogram of the peak positions in (e). (g) $100~{\rm s}$ segment of PLE spectra on NV2 obtained by rapidly scanning a voltage, $V_{AC}$, with feedback applied to $V_{DC}$. %The laser frequency was constant to within $20~{\rm MHz}$ throughout.
    }
\end{figure}
%\end{figure*}

To realize even higher spectral resolution, we performed photoluminescence excitation (PLE) spectroscopy.  Attenuated light ($\sim60~{\rm nW}$) from a tunable, external-cavity diode laser ($\sim637~{\rm nm}$) was used for ZPL excitation near saturation, and the collected light was filtered to direct PSB emission ($650\mbox{-}800~{\rm nm}$) to a single-photon-counting detector. Microwaves resonant with the ground-state spin transition, $D_{GS}=2.877~{\rm GHz}$, were continuously applied to counteract optical pumping \cite{JEL2002,TAM2008}, and light from a repump laser ($532~{\rm nm}$) was occasionally employed to reverse photoionization \cite{DRA1999,SAN2006,WAL2011}.

Figure \ref{fig:feedback}(a) plots PLE spectra for NV1 as a function of $V_a$, applied simultaneously to $V_{\mathrm{ref}}$ and $V_1$, with $V_2=0$. Several excitation lines are resolved due to the presence of resonant microwave excitation. We fit the five strongest lines with Lorentzian profiles. The extracted peak positions are plotted in Fig. \ref{fig:feedback}(b) along with a global fit to the model based on Eq. \eqref{eq:hamiltonian}, yielding Stark coefficients $\Delta d_{\parallel} F_{\parallel}/V_a=0.11(1)~{\rm GHz/V}$ and $d_{\perp} F_{\perp}/V_a=0.26(2)~{\rm GHz/V}$. These coefficients are about four times smaller than those realized under strong $532~{\rm nm}$ excitation, consistent with recent observations of enhanced electrical tuning due to photoionization \cite{BAS2011,BER2012}.

We note that the average linewidth for single scans \cite{FU2009} was $\Gamma_{\mathrm{ss}}=0.14(3)~{\rm GHz}$ for NV1 and $\Gamma_{\mathrm{ss}}=0.48(8)~{\rm GHz}$ for NV2. In both cases, $\Gamma_{\mathrm{ss}}$ is much broader than the natural linewidth, $\Gamma_{\mathrm{nat}}\approx13~{\rm MHz}$, and is independent of scan rate up to $\sim20~{\rm GHz/s}$. This property requires further investigation, as $\Gamma_{\mathrm{ss}}\approx\Gamma_{\mathrm{nat}}$ has been observed elsewhere \cite{TAM2006,FU2009,BER2012}.

%In addition to static control over optical transition frequencies and strengths, electric tuning can be used to provide dynamic compensation for fluctuating local fields. These fluctuations lead to spectral diffusion of the ZPL frequencies \cite{FU2010,ROB2010}, a significant challenge for optical resonance studies \cite{TOG2010,BUC2010,ROB2011,BER2011ARXIV}.
A likely cause for NV spectral diffusion is charge dynamics due to photoionization of nearby defects.%; this effect may be magnified due to the presence of applied voltages \cite{BAS2011}.
To investigate, we use PLE spectroscopy in a different device on the $\ket{0}\rightarrow\ket{E_x}$ transition of a single NV center in natural, type IIa (Ural) diamond. This sample was chosen due to the much narrower linewidth, $\Gamma_{\mathrm{ss}}=60(7)~{\rm MHz}$, even after $\sim15~{\rm MHz}$ of power broadening. Figure \ref{fig:feedback}(c) shows typical PLE spectra for $200~{\rm ms}$ scans with repump pulse ($\sim10~{\rm \mu W}$, $20~{\rm ms}$ duration) applied only after the NV center had photoionized. The transition frequency drifts over a range significantly larger than $\Gamma_{\mathrm{ss}}$ during the $280~{\rm s}$ data set.
%A voltage $V_1=-4~{\rm V}$ was applied ($V_2=V_{ref}=0$), which is thought to magnify spectral diffusion, due to enhanced local charge fluctuations \cite{BAS2011ARXIV,BER2011ARXIV}.

Our solution to the spectral-drift problem is to actively adjust $V_a$ to compensate for the changing local field. We start with $V_a=-4~{\rm V}$, and, during the back-scan of subsequent scans (final $10\%$ of each cycle), we employ software-controlled feedback with the following algorithm. We first determine the position and intensity of the peak fluorescence. If the intensity falls below a threshold, we apply a repump pulse and do not change $V_a$. Otherwise, we change $V_a$ based on optimized proportionality and integration inputs [Supplementary Information].

Figure \ref{fig:feedback}(d) shows PLE spectra under similar conditions as Fig. \ref{fig:feedback}(c) but now with feedback applied. While $\Gamma_{\mathrm{ss}}$ remains unchanged, the center-frequency drift is substantially reduced. We fit the spectra in Figs. \ref{fig:feedback}(c) and (d) with Lorentzian profiles and plot the extracted peak positions in Fig. \ref{fig:feedback}(e). In the case of no feedback, two mechanisms of spectral drift are identified: large instantaneous jumps following application of the repump and slower drift in between repump pulses. Under feedback, spectral jumps still accompany repump pulses, but these are quickly compensated for. Two spectral jumps with the slowest recovery, $5\mbox{-}10$ scans, are shown as insets. A figure of merit for the total drift is obtained by plotting the histogram of fitted peak positions from all PLE scans and determining the resulting standard deviation, $\sigma$ [Fig. \ref{fig:feedback}(f)]. Without feedback, we find a non-uniform profile with $\sigma\approx65~{\rm MHz}$. Under feedback, $\sigma=16~{\rm MHz}$, which is smaller than $\Gamma_{\mathrm{ss}}$ and comparable to $\Gamma_{\mathrm{nat}}$.

This feedback technique can be applied at significantly higher bandwidth (here, up to $20~{\rm Hz}$ scan repetition rate) without compromising stability. Throughout, we find that feedback reduces $\sigma$ to a fraction of $\Gamma_{\mathrm{ss}}$. Similar results were obtained for NV1 and NV2 [Supplemental Information] as well as for stabilizing the $\ket{\pm1}\rightarrow{A_2}$ transition.

It is often advantageous to perform experiments with the excitation laser frequency fixed to an external reference. In this case, voltage feedback can still be employed by sweeping the ZPL transition frequency using an AC voltage, $V_{AC}$, and providing stabilizing feedback to the DC component, $V_{DC}$. With this technique, feedback can be applied continuously without substantially degrading photon indistinguishability, provided that the modulation depth and laser linewidth are sufficiently small.

Figure \ref{fig:feedback}(g) shows results of locking the NV2 $\ket{0}\rightarrow\ket{E_x}$ transition frequency using only applied voltages. We perform PLE spectroscopy as before except, instead of scanning the laser frequency, we ramp the voltage, $V_{AC}$ (applied to $V_2$ and $V_{\mathrm{ref}}$), with amplitude $3~{\rm V_{pp}}$ and period $0.1~{\rm s}$. Meanwhile, $V_{DC}$ is fed back to $V_1$, initially starting at $-8~{\rm V}$, but varying by $\sim\pm4~{\rm V}$ throughout the $600~{\rm s}$ measurement. After background subtraction, we collect on average $144~{\rm cts/s}$. This compares favorably to the $34~{\rm cts/s}$ collected without feedback (with repump applied every scan). The overall count rate can be further increased with improved collection efficiency \cite{HAD2010,HAU2012} and resonant Purcell enhancement \cite{FAR2011,FAR2012ARXIV}.

In summary, we have used the Stark effect to electrically tune and stabilize the structure of the NV center's excited state. Applied simultaneously to a pair of NV centers, these techniques pave the way for increased two-photon interference visibilities \cite{PAT2010,LET2010,BER2012} and heralded entanglement success probabilities \cite{MOE2007}. %

%In order to completely eliminate inhomogenous broadening, further work is necessary to understand and control more rapid broadening mechanisms \cite{SAL2010}.
%Dynamic electric control may also be used to encode spatial information into arrays of NV centers for nanoscale magnetic resonance imaging.
%\end{document}

We acknowledge support by the Defense Advanced Research Projects Agency (award no. HR0011-09-1-0006), the Regents of the University of California, and the Australian Research Council (ARC) (Project Nos. LP100100524, DP1096288, and DP0880466). We thank T. Karle, B. Gibson, T. Ishikawa, B. Buckley, and A. Falk for valuable discussions.

%\bibliography{bibtex}
%

\clearpage

\begin{widetext}
\begin{center}
\textbf{\large Supplementary Information: Dynamic stabilization of the optical resonances of single nitrogen-vacancy centers in diamond}
\end{center}
\end{widetext}

\setcounter{figure}{0}
\setcounter{equation}{0}
\setcounter{page}{1}
\makeatletter
\renewcommand{\thefigure}{S\@arabic\c@figure}
\renewcommand{\theequation}{S\@arabic\c@equation}
\renewcommand{\bibnumfmt}[1]{[S#1]}
\renewcommand{\citenumfont}[1]{S#1}

\section{Electrostatic modeling}
We modeled the electric field produced by the electrode structure in Fig. 1(c) of the main text using the COMSOL MULTIPHYSICS\textregistered~ electrostatics package. The bottom layer of each electrode is composed of 10 nm of Ti, so for simplicity we model the electrodes as being 100-nm thick, composed entirely of Ti. We use a relative permittivity for the diamond substrate $\epsilon_r=5.1$. The positions of NV1 and NV2 were determined by fluorescence micrographs, as in Fig. 1(d) of the main text.

\begin{figure}
\centering
    \includegraphics[width=.4\textwidth]{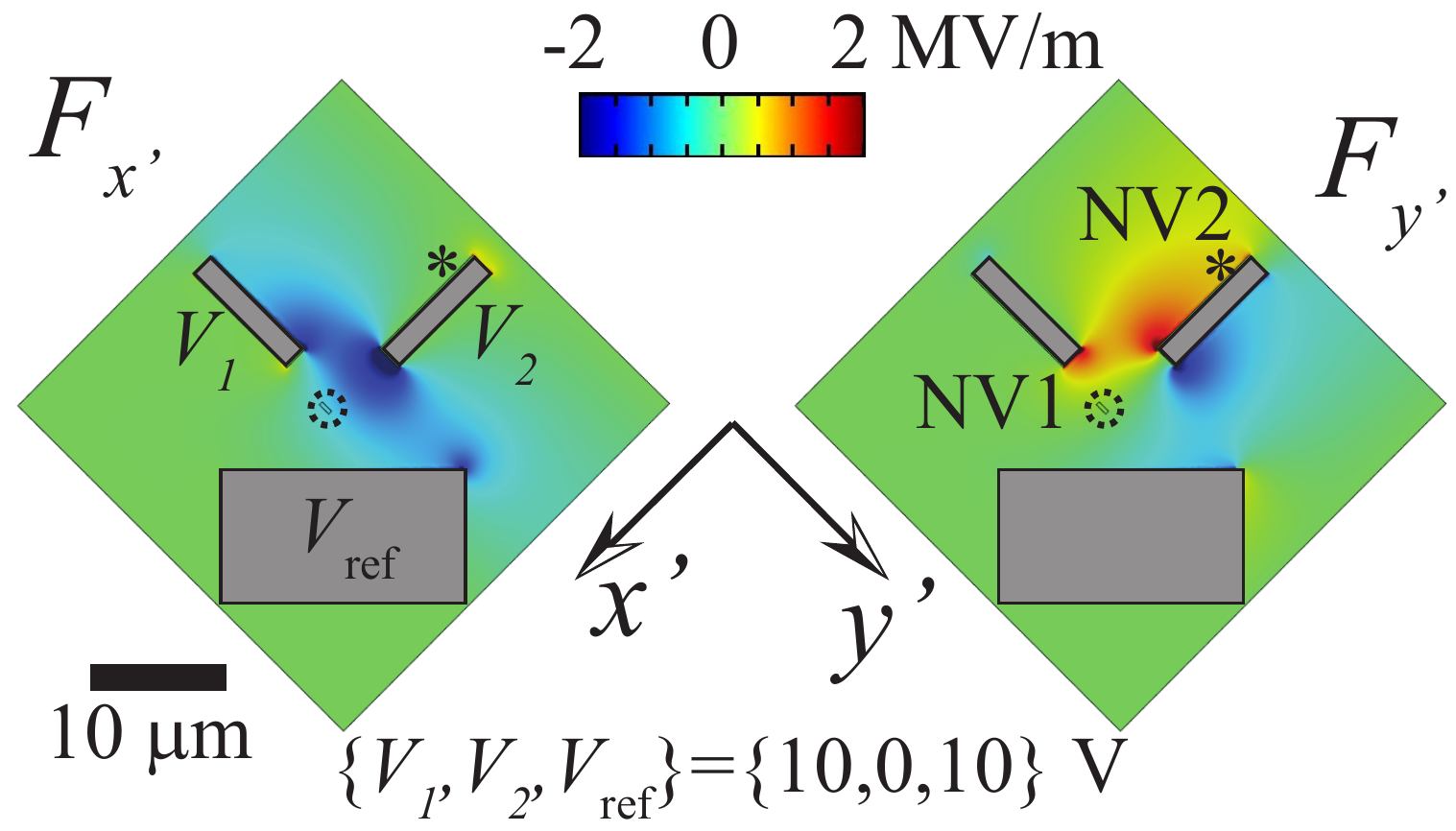}
    \caption{\label{fig:comsol} In-plane electric-field components on the diamond surface, $F_{x',y'}$, for $V_a=10~{\rm V}$ applied simultaneously to $V_1$ and $V_{\mathrm{ref}}$, with $V_2=0$. The locations of NV1 (dashed circle) and NV2 (asterisk) are shown, as in Fig. 1 of the main text. The out-of-plane component $F_{z'}\approx0$.}
\end{figure}

Based on our simulations, application of $10~{\rm V}$ to one electrode (with the other two electrodes grounded) corresponds to an electric field amplitude at the location of NV1 of $0.9, 0.6,$ and $1.1~{\rm MV/m}$, for $V_1, V_2,$ and $V_{\mathrm{ref}}$, respectively. In Fig. \ref{fig:comsol} we plot the in-plane electric field components on the diamond surface for $\{V_1,V_2,V_{\mathrm{ref}}\}=\{10,0,10\}~{\rm V}$.

As a note of caution: this model assumes a perfect dielectric response. In reality the local field is subject to significant deviations due to charge variations introduced by the electrodes or from photo-ionization \cite{BAS2011S}. These deviations do not substantially affect the performance of dynamic ZPL stabilization, since the proportional gain can be adjusted to compensate (see below), but they do play an important role in static tuning. One example is the photo-induced effect that is responsible for the $\sim4$ times greater Stark tuning coefficients for strong ($2~{\rm mW}$) cw green excitation, compared with weak ($\sim60~{\rm nW}$) red excitation, and is discussed in detail in Ref. \cite{BAS2011S}.

We also observe hysteretic charging effects which alter the static tuning even with very weak $(\lesssim60~{\rm nW}$) red excitation and without applying any green repump. This hysteresis results in a variation of observed tuning coefficients which depends on scan direction, rate, and history. The variation exists even between successive voltage scans as short as 10 s. Overall, the observed tuning coefficients vary by up to a factor of 5, making a measurement of the intrinsic dipole moment of the NV center a difficult task. Using the Stark tuning coefficients for NV1, determined from the $\sim2~{\rm hr}$ forward voltage scan in Fig. 3(a) of the main text, the dipole moment is $\{\Delta d_{\parallel},d_{\perp}\}\approx\{4,5\}~{\rm GHz/MV/m}$, which is probably correct to within a factor of 2-3. For this calculation the magnitude of the applied electric field and its angle with respect to the NV axis were estimated based on the electrostatic modeling in Fig. \ref{fig:comsol} and the dependence of NV1 fluorescence intensity on excitation light polarization.

\section{Reproducibility of emission spectroscopy}

\begin{figure*}
\centering
    \includegraphics[width=.97\textwidth]{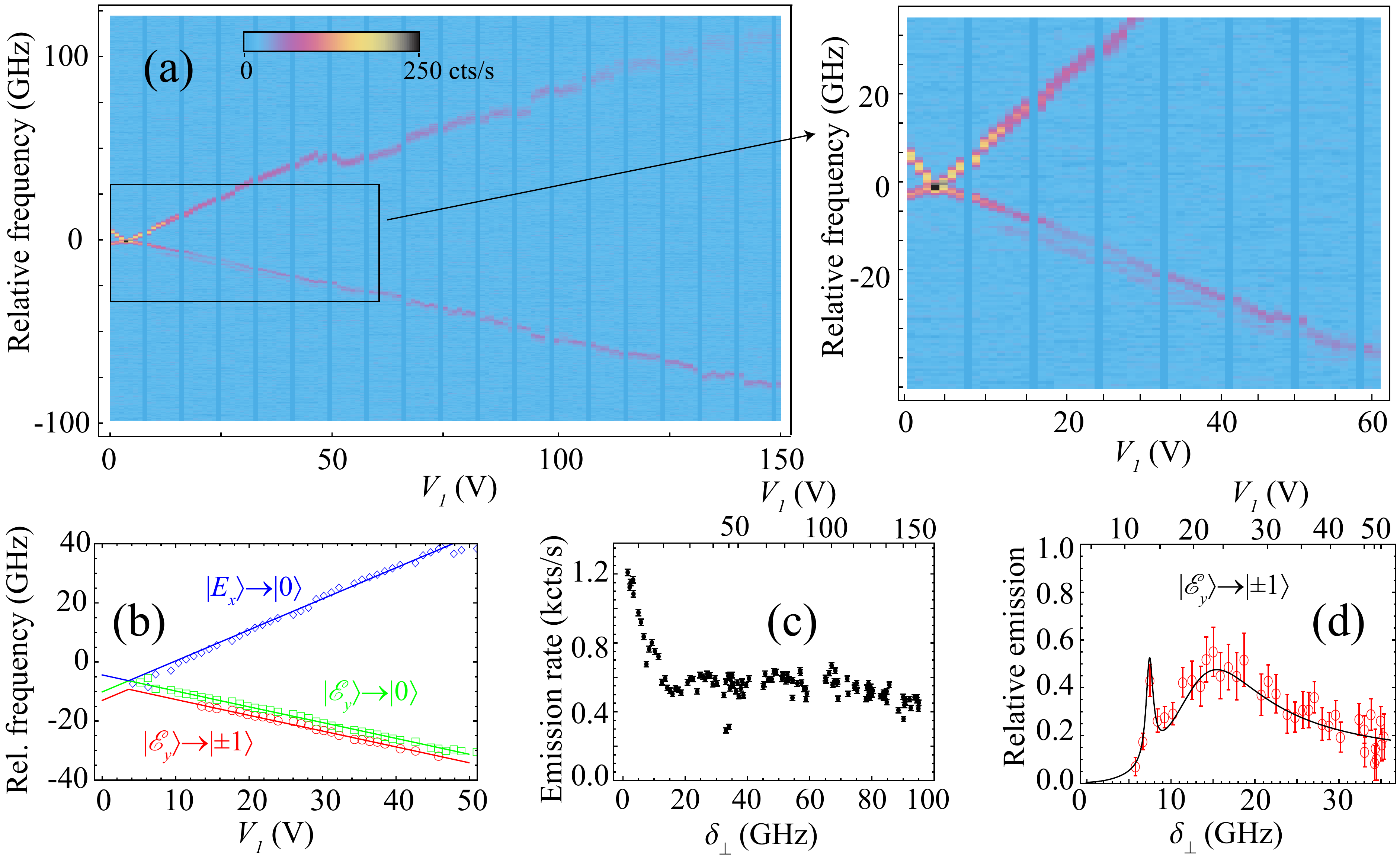}
    \caption{\label{fig:emission_rep} (a) Stark emission spectra as a function of applied voltage on $V_1$, with $V_2=V_{\mathrm{ref}}=0$. Low-field data are displayed on the right. (b) Low-field peak positions determined from Lorentzian fits (symbols) and global fit. Lorentzian fit uncertainty is smaller than the plotted symbols. (c) Total ZPL emission versus $\delta_{\perp}$. (d) Relative intensity of the $\ket{\mathscr{E}_y}\rightarrow\ket{\pm1}$ emission line along with fit. The methods for fitting and processing spectra are described in the main text.}
\end{figure*}

The reproducibility of Stark emission spectroscopy results was tested by varying the voltage on various combinations of electrodes. Figure \ref{fig:emission_rep}(a) shows emission spectra while varying the voltage applied to $V_1$, with $V_2=V_{\mathrm{ref}}=0$. Despite the continuous voltage tuning, small kinks in the emission lines are observed, particularly following dark frames. During the dark frames of this particular data set, we performed a peak-finding procedure that locates the optical focal position which produces maximum NV emission. The resulting shifts in optical focus may produce small changes in the local field due to photo-induced charge redistribution \cite{BAS2011S,BER2012S}. In Fig. 2 of the main text we used a different procedure to maintain optical focus based on continuous feedback using a weak white-light reflection image. This may explain the absence of sharp kinks in the Stark emission spectra in Fig. 2(a).

The inset of Fig. \ref{fig:emission_rep}(a) shows the low-field spectra, with three emission lines clearly visible. As in Fig. 2 of the main text, we do not observe $\ket{A_2}\rightarrow\ket{\pm1}$ emission, indicating ground-state spin polarization, $\mathscr{P}_{GS}\gtrsim85\%$. Following the procedure outlined in the main text, we fit these spectra [Fig. \ref{fig:emission_rep}(b)] and found Stark coefficients of $\Delta d_{\parallel} F_{\parallel}/V_a=0.26(2)~{\rm GHz/V}$ and $d_{\perp} F_{\perp}/V_a=0.81(4)~{\rm GHz/V}$.  In Fig. \ref{fig:emission_rep}(c), the total emission intensity as a function of $\delta_{\perp}$ is plotted. These data are qualitatively similar to that found in Fig. 2(d) of the main text. In Fig. \ref{fig:emission_rep}(d), the emission intensity of the $\ket{\mathscr{E}_y}\rightarrow\ket{\pm1}$ line, relative to the total emission from $\ket{\mathscr{E}_y}$, is plotted versus $\delta_{\perp}$. Based on the fit, we find $\theta_r=12(5)^{\circ}$.

\section{feedback optimization}

\begin{figure}
\centering
    \includegraphics[width=.47\textwidth]{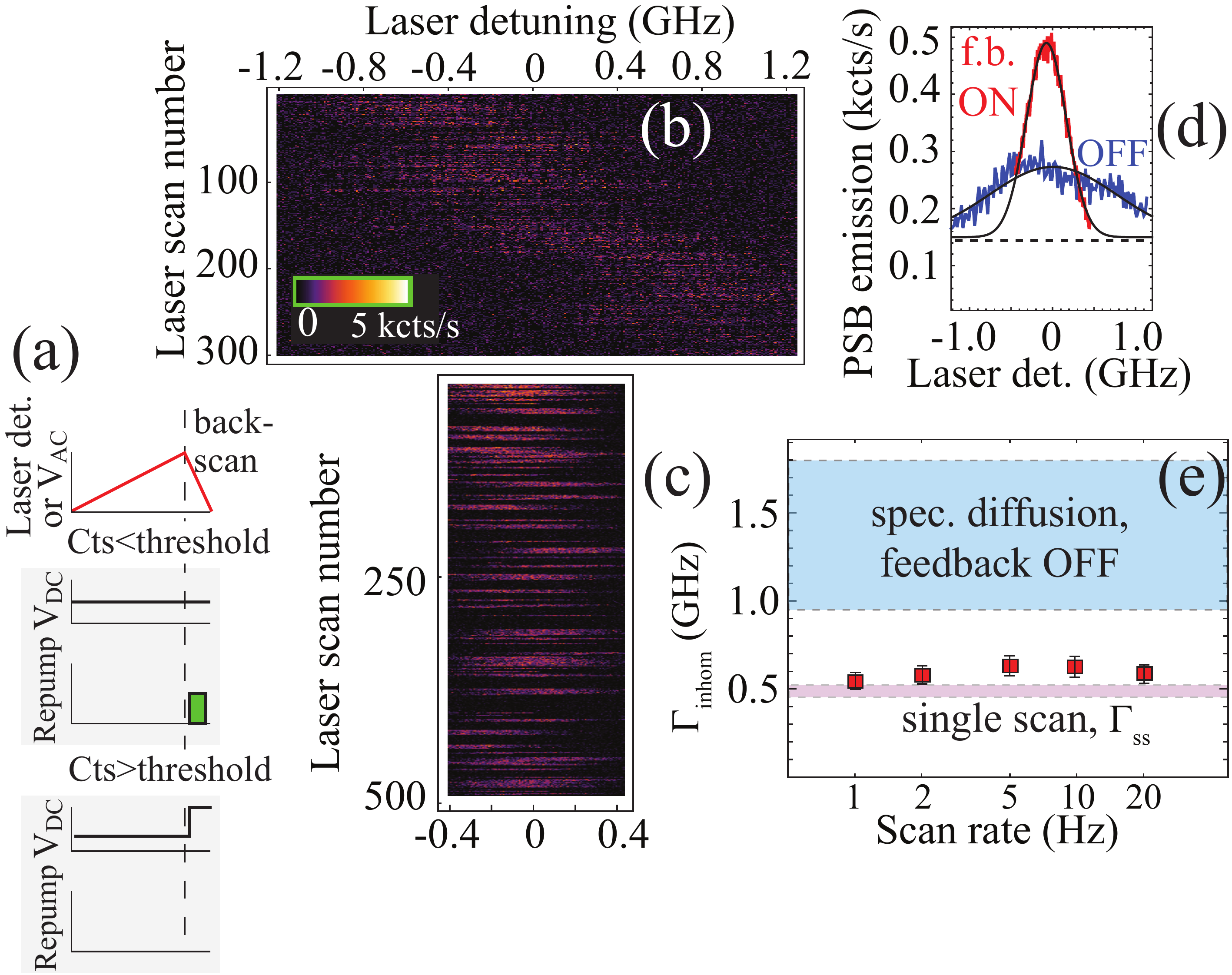}
    \caption{\label{fig:Suppfeedback} (a) Timing diagram of the feedback protocol described in the text. (b) PLE spectra for NV2 ($\ket{0}\rightarrow\ket{E_x}$) with $V_a=-4~{\rm V}$. The scan rate was $1~{\rm Hz}$, and a repump pulse ($10~{\rm \mu W}$, $0.1~{\rm s}$) was applied during every back-scan. (c) PLE spectra for NV2 with feedback applied. The enhanced spectral stability allowed us to reduce the laser-frequency scan range by a factor of four. (d) Sum over the scans in (b) and (c) with, respectively, feedback off (blue) and on (red) along with Gaussian fits (solid black lines) using the mean background of $150~{\rm cts/s}$ (black dashed line). (e) Comparison of $\Gamma_{\mathrm{inhom}}$ as a function of scan repetition rate. The spectral-diffusion-broadened linewidth (blue shaded region) is the range observed over data sets taken without feedback with 1, 2, 5, 10 and $20~{\rm Hz}$ scan rates. %The laser frequency was constant to within $20~{\rm MHz}$ throughout.
    }
\end{figure}

The basic feedback protocol is described in the text, and here we outline experimental details. All PLE scans used a ramp waveform with $90\%$ duty cycle. In other words, we scan either the laser frequency or $V_{AC}$ in one direction for $90\%$ of the total scan cycle and the final $10\%$ is devoted to scanning back in the other direction (the ``back-scan''). We divide our collected PSB counts into bins of variable size, typically forming $10\mbox{-}50$ bins in total. During the backscan of each cycle (denoted with index $i$), we search for the bin location, $b_i$, with the maximum counts, $C_i$. We set a threshold, $T$, typically corresponding to a count rate of $1~{\rm kcts/s}$, much larger than the background signal off resonance. If $C_i<T$, we apply a green repump pulse for the remainder of the backscan and do not change $V_{DC}$.

If $C_i\geq T$, we do not apply a repump pulse. Instead, we change $V_{DC}$ by an amount $\delta V_i$ using the following formula:
\begin{equation}
\label{eq:feed}
\delta V_i = G\times(B-\frac{1}{N}\sum_{j=0}^{N-1} b_{i-j}).
\end{equation}
Here $G$ is a gain factor, $N$ is an integration factor corresponding to the number of cycles used to determine $\delta V_i$, and $B$ is the desired peak position. In our experiments, we typically set $B$ to be the bin at the center of each scan. For all of the laser-frequency scans in the main text [Fig. 3(d)-(f)], we used $N=1$ (no integration). For the $0.1$-s voltage scans we found that feedback was most efficient using $N=2\mbox{-}4$. The $100$-second portion shown in Fig. 3(g) used $N=2$. We separately optimize $G$ based on the NV center's Stark tuning coefficients as well as the method and rate of scanning. Figure \ref{fig:Suppfeedback}(a) depicts a timing diagram of the feedback routine.

We performed the feedback routine discussed above on the $\ket{0}\rightarrow\ket{E_x}$ transition of NV2 in the NV-doped surface layer sample. Figure \ref{fig:Suppfeedback}(b) shows typical PLE spectra when scanning the excitation laser frequency through resonance and applying a repump after each scan. As mentioned in the main text, the average linewidth for single scans, computed using the technique in Ref. \cite{FU2009S}, was $\Gamma_{\mathrm{ss}}=0.48(8)~{\rm GHz}$ for NV2. Nonetheless, the ZPL center frequency in Fig. \ref{fig:Suppfeedback}(c) drifts over a range significantly larger than $\Gamma_{\mathrm{ss}}$ during the $300~{\rm s}$ data set.
%A voltage $V_1=-4~{\rm V}$ was applied ($V_2=V_{ref}=0$), which is thought to magnify spectral diffusion, due to enhanced local charge fluctuations \cite{BAS2011ARXIV,BER2011ARXIV}.

Figure \ref{fig:Suppfeedback}(c) shows PLE spectra under similar conditions as Fig. \ref{fig:Suppfeedback}(c) but now with feedback applied. While $\Gamma_{\mathrm{ss}}$ remains unchanged, the center-frequency drift is substantially reduced. A figure of merit for the spectral drift of the transition is obtained by summing over many PLE scans and determining the resulting inhomogenous linewidth, $\Gamma_{\mathrm{inhom}}$. This figure-of-merit is somewhat different from the histogram technique described in the main text and was employed due to the large single-scan linewidth for this NV center. Figure \ref{fig:Suppfeedback}(d) compares the sum of spectra with feedback off [Fig. \ref{fig:Suppfeedback}(b)] and on [\ref{fig:Suppfeedback}(c)] along with Gaussian fits. Without feedback, we find $\Gamma_{\mathrm{inhom}}=1.8(2)~{\rm GHz}\approx3.8\Gamma_{\mathrm{ss}}$, and with feedback this decreases to $0.54(4)~{\rm GHz}\approx1.1\Gamma_{\mathrm{ss}}$.

This feedback technique can be applied at significantly higher bandwidth without compromising stability. Figure \ref{fig:Suppfeedback}(e) plots $\Gamma_{\mathrm{inhom}}$ as a function of scan repetition rate up to $20~{\rm Hz}$. Throughout this range, we find $\Gamma_{\mathrm{inhom}}\lesssim1.3\Gamma_{\mathrm{ss}}$. Future improvements could involve ultra-fast correlation measurements to determine the nature of the broad single-scan linewidth \cite{SAL2010S}.

\end{document}